\documentstyle[epsfig,aps,preprint]{revtex}

\begin{document}
\tolerance=100000
\author{ Zafar Ahmed \\
Nuclear Physics Division, Bhabha Atomic Research Centre \\
Trombay, Bombay 400 085, India \\ zahmed@apsara.barc.ernet.in }
\title
{Gaussian-random Ensembles of Pseudo-Hermitian Matrices \footnote {Invited Talk
delivered at II International Workshop on `Pseuo-Hermitian Hamiltonians in
Physics'at Prague, June 14-16, 2004.}}
\date{\today}
\maketitle
\begin{abstract}
Attention has been brought to the possibility that statistical fluctuation
properties of several complex spectra, or, well-known number sequences may display strong
signatures that the Hamiltonian yielding them as eigenvalues is PT-symmetric (Pseudo-Hermitian).
We find that the random matrix theory of pseudo-Hermitian
Hamiltonians gives rise to new universalities of level-spacing distributions
other than those of GOE, GUE and GSE of Wigner and Dyson. We call the new proposals
as Gaussian Pseudo-Orthogonal Ensemble and Gaussian Pseudo-Unitary Ensemble.
We are also led to speculate that the enigmatic Riemann-zeros (${1 \over 2}\pm i t_n)$
would rather correspond to some PT-symmetric (pseudo-Hermitian) Hamiltonian.
\end{abstract}

\section{Introduction}
\par A large body of spectra of the bound levels and resonances is available in
nuclear physics wherein the nuclear interaction Hamiltonian is unknown.
The well-known prime numbers 2,3,5,7,9,11,13,17,19,... do not have a
representation so far. One may wonder if there is a Hermitian Hamiltonian
that can yield them as its discrete eigenvalues. The zeta function, $\zeta(z)$,
which is real on real line as per the hypothesis of Riemann (1859) has {\it all}
its non-real zeros as ${1 \over 2}\pm it_n$ ($t_n$ being real) [1]. The marathon and accurate
computation of more than $10^{20}$ zeros of $\zeta(z)$ by Odlyzko [2] testifies RH the best, with not
even a single
exception so far. First few zeta-zeros (Riemann-zeros : RZs) are given by $t_1=14.13,
t_2=21.02,t_3=30.42,t_4=37.58.$
Hilbert and Polya have conjectured that $t_n$ could be like the eigenvalues of a Hermitian
Hamiltonian.
Consequently the completeness of the spectra will lead to the proof of one of
the most enigmatic and formidable problems called RH.
\par In nuclear spectroscopy, the pressing need was to identify the universality
that underlies the nuclear levels. The Nearest Neighbour Level Spacing Distribution
(NNLDS) is the statistical distribution of the fluctuations around the mean level spacing
of a collection of spectra under a class of fixed parity or other quantum numbers.
Random Matrix Theory (RMT)
was discovered in the late 1950s to predict universalities of NNLSD in various situations.
Among the notable names we have Wigner, Landau, Dyson, Gaudin, Mehta, Porter,
Ginibre and Pandey to associate with RMT [3].
\par Since the interaction Hamiltonian is not known it would rather be taken
as non-integrable, then there are three types of NNLSDs obtained by Wigner and
Dyson. These are given as in Eq. (4) (see below) and called as GOE,GUE and GSE
spacing-statistics [3].
Most remarkably the nuclear levels with same $(J,\pi)$ are well-known to follow
GOE statistics. The energy levels of the chaotic Sinai-billiard are  known to follow
the same statistics. This has supported the idea that chaos may be there in
nuclear dynamics too. Sinai-billiard refers to a particle in a square region with
hard, reflecting edges along with a hard, reflecting circle in its center.
\par Well before the advent of RMT, Hilbert had prophesied that the real part of
RZ are distributed as the eigenvalues of certain random Hermitian matrices.
In fact, in terms of RMT it means that the real parts of RZs obey NNLSD corresponding
to GUE. Over a hundred years old, this prophecy  of Hilbert has been testified by Odlyzko
[2] as late as 1989 using more than $10^{20}$ RZs. Montgomery in 1973 (see in Mehta [3])
analytically derived
two-point correlation function of the real parts of RZs which turned out
to be the same as that of GUE. Dyson had already expected this and it has also been
confirmed numerically by Odlyzko [2]. These two remarkable affirmations have strengthened
the Hilbert-Polya conjecture to look for a Hermitian Hamiltonian for the RH.
\par Last few years have witnessed an interesting phenomenon whereby the real discrete
eigenvalues need not necessarily be possessed by Hermitian Hamiltonians. Non-Hermitian,
PT-symmetric [4] or pseudo-Hermitian [5,6] Hamiltonians too can possess real discrete
spectrum.
In RMT, the matrix ensembles GOE, GUE, GSE refer to Hamiltonians with TRI (Time
Reversal Invariance), without TRI, and with TRI including Kramer's
degeneracy respectively. Recently, we have developed the Gaussian-random ensembles of
pseudo-Hermitian matrices, that give rise to new ``universalities" of NNLSD.
We have called the new ensembles as GPUE [7,8] which are expected to represent
the cases where Parity (P) and Time-reversal (T) symmetries are individually
broken but preserved jointly.
\par In this paper, we would like to present GPUE with more refinements
and reorientations. In Section 2, we briefly introduce RMT with the ensembles
of Wigner and Dyson. We then find a natural scope to go in for new ensembles.
In Section 3 and 4 the new ensembles are described. In Section 5, we report
two interesting dichotomies where spacing statistics are like GOE and GUE,
despite Hamiltonians being pseudo-Hermitian. Following one of these dichotomies,
in Section 6, we speculate on the possible features of Hamiltonian corresponding to RH.
We present a summary of conclusions in Section 6.
\section {Wigner-Dyson Ensembles of Gaussian-Random Matrices : GOE, GUE, GSE}
An eigenspectrum is (practically) an outcome of the diagonalization of a Hamiltonian
matrix. Since the analysis of NNLS requires at least two eigenlevels, in RMT, one
begins with $2 \times 2$ Hamiltonian matrix for simplicity. The RMT takes an
important note of the fact that for systems with TRI, the matrix
Hamiltonians are real-symmetric $(H^R)$; for systems without TRI, the Hamiltonians are
Hermitian matrices $(H^H)$ and for systems with TRI plus Kramer's degeneracy, the Hamiltonians are
even (at least 4) dimensional $(H^K)$.  These Hamiltonian matrices are given as [3,9] :
\begin{eqnarray}
H^R =\left [\begin {array} {cc} a+b & c \\ c & a-b \end {array} \right],
~~H^H =\left [\begin {array} {cc} \alpha  & \gamma \\ \gamma^\ast & \beta \end {array} \right],
H^K =\left [\begin {array} {cccc} \alpha & 0 & \gamma^\ast & -\delta \\
0 & \alpha & \delta^\ast & \gamma\\
\gamma & \delta & \beta & 0\\ -\delta^\ast & \gamma^\ast & 0 & \beta \\
\end {array} \right],
\end{eqnarray}
where $\alpha=a+b,\beta=a-b,\gamma=c+id,\delta=e+if$.
In order to have a large collection of eigenvalue pairs, we assume the entries
$a,b,c,d,e,f$ are real and drawn independently from a Gaussian-random population.
Notice that respective eigenvalues for (1) are
\begin{equation}
E^R_{1,2}=a\pm\sqrt{b^2+c^2},~~ E^H_{1,2}=a\pm \sqrt{b^2+c^2+d^2},~~ E^K_{1,2}=a\pm
\sqrt{b^2+c^2+d^2+e^2+f^2}.
\end{equation}
In RMT, to reemphasize, the energy eigenvalues for TRI systems turn out to
have the form as $E^R$, for non-TRI systems they are as $E^H$ and for TRI systems
with Kramer's degeneracy  are as $E^K$s.
The level-spacings ($s=|E_1-E_2|$) in various cases are
\begin{equation}
s^R\sim \sqrt{b^2+c^2},~~ s^H\sim \sqrt{b^2+c^2+d^2},~~s^K \sim
\sqrt{b^2+c^2+d^2+e^2+f^2}.
\end{equation}
Now the question to be asked is : what is the probability distribution of the
level-spacing $s$, when $a,b,...,f$ are Gaussian-random variables?
More importantly one wants to know whether the levels have tendency to repel
or attract each other and then what the degree of repulsion/attraction is.
The real-symmetric matrices have an Orthogonal symmetry which corresponds to SO(N)
: GOE. The Hermitian matrices have unitary symmetry which corresponds to SU(N)
: GUE. The matrices $H^K$ have Symplectic symmetry corresponding to Symplectic
group $S_p(N)$ : GSE. These facts are used to derive Wigner-Dyson universalities in NNLSD as
[3,9]
\begin{equation}
P^{GOE}(x)= {\pi \over 2}x e^{-{\pi x^2 \over 4}},~~P^{GUE}(x)={32 \over \pi^2} x^2
e^{-{4x^2 \over \pi}},~~P^{GSE}(x)={2^{18} \over 3^6 \pi^3} x^4 e^{-{64 \over 9 \pi}x^2}.
\end{equation}
Here $x$ is a scaled spacing ($s$) with respect to its mean value
($\langle s \rangle$). For smaller values of spacing $s$, notice the tendency of level
repulsion as $P(x \rightarrow 0)\rightarrow 0$. The degree of repulsion is linear, quadratic and quartic
respectively for GOE, GUE and GSE. Most importantly these distributions (4)
turn out to be excellent approximants to $P(x)$ for $N\times N$
matrices [3,9].
\par An alternative intuitive way of looking into these NNLSD is to see that
the three statistics (4) correspond to the following multidimensional integral
\begin{equation}
P^{GOE}(s)~\sim~ \int_{-\infty}^{\infty} \int_{-\infty}^{\infty} \int_{-\infty}^{\infty}
e^{-(a^2+b^2+c^2)}~ \delta(s-\sqrt{b^2+c^2})~ da~ db~ dc,
\end{equation}
and similarly others for GOE and GSE. \\
The common feature of the eigenvalues (2) or spacings (3) lies in their
absolute reality. Let us ask the following questions : Firstly, can there be Hamiltonians possessing
conditionally real eigenvalues or spacings e.g., $s\sim \sqrt{b^2-c^2}$
(real iff $b^2 \ge c^2$ )and $s\sim \sqrt{b^2+c^2-d^2}$ (real iff $b^2+c^2 \ge d^2$)?
Secondly, what are the symmetries of such Hamiltonians? Thirdly, what are new
universalities of NNLSD ? Such questions have led us to think of new Gaussian-random
ensembles of pseudo-Hermitian Hamiltonians [7,8]. By a Gaussian ensemble, we would mean that
the probability distribution of Hamiltonian $H$ is commonly given as
\begin{equation}
P(H)={\cal N}~e^{-{Tr(HH^\dagger)\over 2\sigma^2}}.
\end{equation}
\section {Gaussian Pseudo-Orthogonal Ensembles (GPOE)}
\begin{center}
{\bf Pseudo-symmetric or complex-symmetric matrix Hamiltonians}
\end{center}
Let us consider a matrix Hamiltonian H given below which is pseudo-Hermitian
as $\eta H {\eta}^{-1}=H^\dagger$ [5] and pseudo-real i.e.,
$\rho H {\rho}^{-1}=H^\ast [11].$
It is self-pseudo-adjoint or symmetric as $H^\prime=H$.
Here $\eta$ are $\rho$ are preferably involutary operators.
It has got conditionally
real eigenvalues iff $b^2 \ge c^2$ Here prime, asterisk (${\cal K}_0$) and dagger
denote transpose, conjugate and transpose-conjugate, respectively.
\begin{eqnarray}
H =\left [\begin {array} {cc} a+b & i c \\ i c & a-b \end {array} \right],~~b^2 \ge c^2,~~
\eta =\rho= \left [\begin {array} {cc} 1  & 0 \\ 0 & -1 \end {array} \right],
~~E_{1,2}=a\pm \sqrt{b^2-c^2}.
\end{eqnarray}
One can construct an antilinear commutant $\Theta = \rho^{-1} {\cal K}_0$ [10]
of $H$ such that $[\Theta, H]=0$ or $ \Theta H {\Theta}^{-1}=H$ and $\Theta^2=1$. We would
like to assert that here $P=\rho^{-1}$ and $T={\cal K}_0$ and hence the antilinear symmetry
$\Theta=PT$. When eigenvalues are real ($b^2 >c^2$), we have $PT \Psi_n= (-1)^n \Psi_n$.
When $b^2 <c^2$, the PT-symmetry is spontaneously broken. This Hamiltonian in
our opinion is another realization of Hamiltonians with antilinear symmetry as
visualized by Haake (page 217 in [9]) as $[{\cal A}, {\cal D}]= 0 $ such that
${\cal A}^2=1$.
\par We define pseudo-orthogonal transformation as
$~_pO^\prime=\delta ~_pO^{-1} \delta^{-1}$ such that for any two arbitrary vectors
from a linear space the scalar product remains invariant i.e.,  ${\tilde x}^\prime
\delta {\tilde y}= x^\prime \delta y$, where ${\tilde x}= ~_pO x$  and
${\tilde y}= ~_pO y$. Let us represent $~_pO$, energy-eigenvalue matrix $E$
and a metric $\delta$
\begin{eqnarray}
~_pO =\left [\begin {array} {cc} \cosh \theta & i \sinh \theta \\ -i \sinh \theta
& \cosh \theta \end {array} \right],
~~\delta = \left [\begin {array} {cc} 0  & -i \\ i & 0 \end {array} \right]=\sigma_y,
~~E = \left [\begin {array} {cc} E_1  & 0 \\ 0 & E_2 \end {array} \right].
\end{eqnarray}
where $-\infty < \theta < \infty$. The single parameter matrix $~_pO$, is expressible as
$\exp(2i\theta J_2)$ with $J_2={1 \over
 2}i\sigma_y$, constitutes a subgroup of $SU(1,1)$ [12]. A very important consequence of the
group connection is that we can generate {\it all} possible $H$ in (7) as
$H=~_p]O~ E ~_pO^{-1}$. This provides us with a unique connection between $(a,b,c)$ and
$(E_1,E_2,\theta)$ and the consequent Jacobian is ${\cal J}={|s| \over 8}$. We have
\begin{equation}
a={E_1+E_2 \over 2},~~b={E_1-E_2 \over 2} \cosh 2\theta,~~ c=-{E_1-E_2 \over 2}
\sinh 2\theta.
\end{equation}
For brevity, we have used $s=E_1-E_2$ and $t=E_1+E_2.$ We can write the probability
distribution, $P(H)$ (6) for the Hamiltonian in (7) as
\begin{equation}
P(a,b,c)={\cal N} e^{-(a^2+b^2+c^2) \over \sigma^2}.
\end{equation}
Using (9) we can transform (10) in terms of $(t,s,\theta)$, further integration over $\theta$
on $[-\infty,\infty]$ gives Joint Probability Distribution function of $(E_1,E_2)$   as
\begin{equation}
P(E_1,E_2)={\cal N}^\prime s K_0\left ( {s^2 \over
4\sigma^2}\right ) e^{-{t^2 \over 4\sigma^2}}.
\end{equation}
Next the integration over $t$ on $[-\infty,\infty]$ yields the NNLSD as
\begin{equation}
P(s)={\cal N}^{\prime\prime}~s~K_0\left({s^2 \over 2\sigma^2}\right).
\end{equation}
By finding $\langle s \rangle$ using (12) and introducing $x={s \over
\langle s \rangle}$, we eventually find
the normalized NNLSD and call it as $P^{GPOE}(x)$
\begin{equation}
P^{GPOE}(x)={\Gamma^4(-{1 \over 4}) \over 32\pi^3}~x~K_0 \left({2\Gamma^4(3/4)
\over \pi^2} x^2 \right).
\end{equation}
If we write it as $P^{GPOE}(x)=\alpha ~x~ K_0(\beta x^2)$, we have $\alpha=0.5818$
and $\beta=0.4569$. When $0 <x<0.5$, we have $P^{GPOE}(x)\sim (0.5-1.2\ln x)x$.
For any other pseudo-symmetric or complex symmetric matrix
Hamiltonian that is composed of three independent Gaussian-random variables
$(a,b,c)$ appearing linearly in $H$, we claim that $P^{GPOE}(x)$ is the universality.
The new ``universality'' shows a distinctly different behaviour as compared to the
usual ones (see Fig. 1(b)).
\section {Gaussain Pseudo-Unitary Ensembles}
\begin{center}
{\bf Pseudo-Hermitian matrix Hamiltonians}
\end{center}
We now consider pseudo-Hermitian matrix Hamiltonians with four parameters
$(a,b,c,d)$
\begin{eqnarray}
H =\left [\begin {array} {cc} a+b & d+i c \\ -d+i c & a-b \end {array} \right],
e^2=b^2-c^2+d^2 \ge 0,\eta= \left [\begin {array} {cc} 1  & 0 \\ 0 & -1
\end {array} \right],E_{1,2}=a\pm e .
\end{eqnarray}
Here we have $\eta H \eta^{-1}=H^\dagger$, $P$ and $T$ operators can be constructed
as prescribed in [10] an and antilinear commutant, $\Theta$, of $H$ can be constructed as
prescribed in [11]. Consider a transformation $~_pU$ which preserves
the pseudo-norm as $\tilde x^\dagger \eta \tilde y= x^\dagger \eta y$, where
$\tilde x=~_pU x, \tilde y=~_pU y$. In doing so $~_pU$ would satisfy an interesting
condition i.e, $~_pU^\dagger = \eta ~_pU^{-1} \eta^{-1}$ which is called pseudo-unitarity.
(see e.g., [7])
\par A general three parameter $(\theta,\psi,\phi)$ matrix, $~_pU$, which is
pseudo-Unitary under the same metric $\eta$ (14) can be written as
\begin{eqnarray}
~_pU =\left [\begin {array} {cc} e^{i\psi} \cosh \theta & e^{i\phi}\sinh \theta\\ e^{i\psi}
\sinh \theta & e^{-i\phi} \cosh \theta \end {array} \right], 0 \le \phi,\psi \le 2\pi,
0< \theta < \infty.
\end{eqnarray}
This constitutes a Lie group $SU(1,1)$ [12] with generators as $J_0={1\over 4}\sigma_z,
J_1={1 \over 4}i\sigma_y, J_3=-{1 \over 4}i\sigma_x$. However, in order to construct
the pseudo-Hermitian matrix (14) we require only two parameters in $~_pU$. The same
situation arises [3,9] in case of GUE, where two out of three parameters
suffice in writing the unitary matrix $U$, nevertheless
it requires three-parameters to have $SU(2)$. Thus, we take $\psi=0$ in (15) and
generate $H$ in Eq. (14) as $~_pU E ~_pU^{-1} =H.$ This is how we go over to $(E_1,E_2,
\phi,\psi)$ from $(a,b,c,d)$. We find
\begin{equation}
a={t\over 2},~ b={s \over 2} \cosh 2\theta,~ c= -{s\over 2} \sinh 2\theta \cos \phi,~
d= {s\over 2} \sinh 2\theta \sin \phi ~ \mbox {and}~ {\cal J}={s^2 \over 4}\sinh 2\theta.
\end{equation}
The probability function (6) for $H$ (14) works to $P(a,b,c,d)={\cal N}
\exp[-(a^2+b^2+c^2+d^2)/\sigma^2]$. A similar procedure (yet more involved)
as done in Section 3, from Eq. (10) to (13), leads us to a new NNLSD as
\begin{equation}
P^{GPUE}(x)= {{\cal B}^2 \over 2(\sqrt{2}-1)}~x~ e^{{\cal B}^2 x^2 \over 4}
~\mbox {erfc} ~\left ({\cal B} x \over \sqrt{2} \right ),~~ {\cal B}={2(\sqrt{2}-\log(1+\sqrt{2}))
 \over \sqrt{\pi} (\sqrt{2}-1)}.
\end{equation}
If we write as $P^{GPUE} (x)= \alpha~ x~ e^{\beta x^2} \mbox {erfc}(\gamma x)$
where $\alpha=2.5433,~\beta=0.5267,~\gamma=1.0263 $. Its linear
dependence on $x$ is deceptive, its behaviour near
small values of $x$ is actually curved (short dashed line in Fig. 1(b))
lying below the curve corresponding to $P^{GPOE}(x)$ (see solid line in
Fig. 1(b)). For $0 <x <0.5$, we have $P^{GPUE}(x) \sim 2.5x(1-.95x).$
\section{Interesting dichotomies and speculations on RH}
\begin{center}
{\bf Quasi-Hermitian matrix Hamiltonians}
\end{center}
Pseudo-Hermitian Hamiltonians under a definite metric as given below are called
quasi-Hermitian. Consider 3- and 4-parameter cases of such $2\times 2$ matrix Hamiltonians.
\begin{eqnarray}
H_3 =\left [\begin {array} {cc} a & (b+ic)/\epsilon \\ (b-ic)\epsilon & a \end {array} \right],
~~H_4 =\left [\begin {array} {cc} \alpha & \gamma/\epsilon \\ \gamma^\ast\epsilon & \beta
\end {array} \right],
~~\eta =\left [\begin {array} {cc} \epsilon  & 0 \\ 0 & 1/\epsilon \end {array} \right].
\end{eqnarray}
Here $\alpha=a+b, \beta=a-b,\gamma=c+id$.
These matrices, despite being pseudo-Hermitian possess absolutely real eigenvalues.
By constructing one and two parameter pseudo-unitary transformation matrices
and by carrying out the procedure outlined in Sections, 3 and 4, the spacing
distributions have been obtained. Interestingly, for $H_3$ the Wigner surmise, $P^{GPOE}(x)$
has been recovered identically [8] ! Even more interestingly, by defining
$\epsilon=e^{-\kappa} $, we find a new analytic expression [13] for $P(x)$ that
hardly differs from $P^{GUE}(x)$ for $\kappa$ from 0  to 0.5. For other
values of $\kappa$ the differences between are also not considerable.
\par For $\epsilon=1$, the Hamiltonian is Hermitian, by changing $\epsilon$
it becomes non-Hermitian such that the spacing distribution does not change
appreciably. We may therefore take such Hamiltonians and interpret them as
a smooth perturbation of the Hermitian Hamiltonian.
\par We feel that the display of $P^{GUE}(x)$ by a certain class of spectra,
though the underlying Hamiltonian is not Hermitian (instead it is quasi-Hermitian)
is a very remarkable result.
\begin{figure}[h]
\psfig{figure=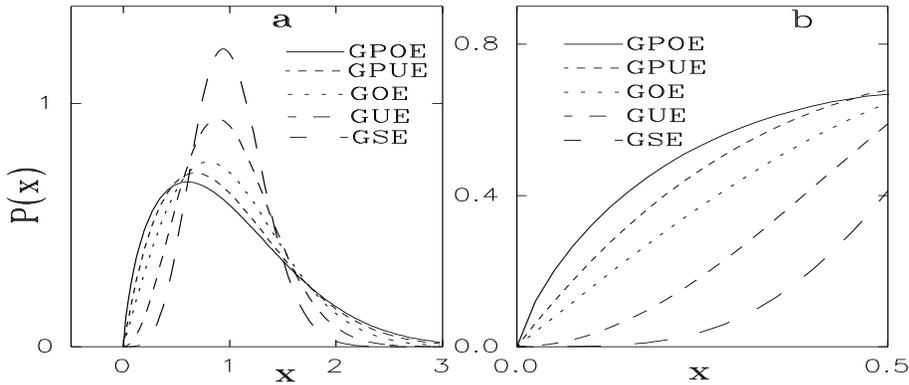,width=12cm,height=5.cm}
\caption {a: Various spacing statistics, $P(x)$, see Eqs. (4,13,17);
b: $P(x)$ for $0<x<0.5$}
\end{figure}
\par Recall that RZs display $P^{GUE}(x)$ so the prospective
Hamiltonian is expected to be both Hermitian and TRI-breaking. In the light
of our dichotomous result, we speculate that the Hamiltonian relevant to RH
could also be a PT-symmetric (Hamiltonian). Since complex-conjugate
eigenvalues are found for a PT-symmetric Hamiltonian when PT-symmetry is
spontaneously broken, so ${1\over 2}\pm it_n$ would naturally follow from a
PT-symmetric Hamiltonian. Furthermore, the vanishing of the norm (PT-norm) of
the eigenstates in the domain where spontaneous breaking of symmetry occurs
can be seen to be directly connected to a crucial criterion proposed by Alain
Connes [1] for a Hamiltonian which could be relevant to RH. Recently,
we have found [13] that for RH a
few Hermitian Hamiltonians proposed so far [14] do not even possess a discrete
spectrum.
\section{Conclusion}
The present work tries to answer the basic question as to what could be the
random matrix theory of currently researched  pseudo-Hermitian Hamiltonians.
The new ``universalities" in Eqs. (13) and (17) as displayed in Fig. 1  are
distinctly different from the usual ones, resulting in weaker repulsion among
the energy-levels. Like the well established GOE,GUE and GSE, here it remains
to be proved that the claimed results for $2\times 2$ matrices would actually stay
on at least as good approximants for the $N\times N$
case. This is an open challenge an answer to this would actually take us from
``universalities'' to universalities.
\par However, we feel that `the weaker level repulsion for small spacings'
is the essence of pseudo-Hermiticity. And since the pseudo-Hermiticity has been
re-cast in terms of more physical PT-symmetry, an observance of weaker level
repulsion at small spacings would suggest an individual violation of P
and T symmetries and joint invariance of PT.
\par PT-symmetric (pseudo-Hermitian) Hamiltonian as a physical model
may now be far, nevertheless signature such as weaker level repulsion would
give way to such Hamiltonians.
\par Two interesting dichotomies have also been presented~-~ owing to one of them
we have speculated that the prospective Hamiltonian for the Riemann Hypothesis would
rather be PT-symmetric (pseudo-Hermitian).
\par Lastly, we would like to point out that in contrast to the Ginibre matrix ensembles
and some more suggested by Haake [9], we consider such complex matrices
(pseudo-Hermitian) which give conditionally real eigenvalues and we define
level spacing as $(E_1-E_2)$ and {\it not as modulus of difference of two complex
eigenvalues}.
\section*{References}
\begin{enumerate}
\item A. Connes, Selecta Math. New Ser. {\bf 5} (1999) 29.
\item A. Odlyzko, AMS Contemporary Math. Series {\bf 290} (2001) 139.
\item M.L. Mehta {\em Random Matrices} (1991) (New-York, Academic); C.E. Porter
{\em Statistical theories of spectra : Fluctuations} (1965) (New-York, Academic).
\item C.M. Bender and S. Boettcher, Phys. Rev. Lett. {\bf 80} (1998) 5243.
\item See e.g.,
A. Mostafazadeh, J. Math. Phys. {\bf 43} (2002) 205;
Z. Ahmed, Phys. Lett. {\bf 290} (2001) 19; {\bf 294} (2001) 287.
\item Proceedings of `$1^{st}$ Intl. Workshop on
{\em Pseudo-Hermitian Hamiltonians in Quantum Physics}' Czech J.
Phys. {\bf 54} (2003); also this volume.
\item Z. Ahmed and S. R. Jain, Phys. Rev. E {\bf 67} (2003) R 045106;
J. Phys. A: Math. Gen. (special issue : Random  Matrix Theory) {\bf 36} (2003)
3349.
\item Z. Ahmed, Phys. Lett. A {\bf 308} (2003) 140.
\item F. Haake, {\em Quantum Signatures of Chaos} (Springer Verlag : N.Y) (1992).
\item C.M. Bender, D.C. Brody, and H.F. Jones, Phys. Lett. {\bf 89} (2003) 270401;
A. Mostafazadeh, J. Math. Phys. {\bf 44} 974; Z. Ahmed, J. Phys. A : Math. Gen.
{\bf 36} 9711; Phys. Lett.  A {\bf 310} 139.
\item Z. Ahmed, J. Phys. A : Math. Gen. {\bf 36} (2003) 10325.
\item N. Mukunda, J. Math. Phys. {\bf 8} (1967) 2210.
\item Z. Ahmed and S. R. Jain, (2004) to be submitted.
\item M.V. Berry and J.P. Keating, SIAM Rev. {\bf 41} (1999) 236;
 Okubo, J. Phys. A : Math. Phys. {\bf 31} (1998) 1049;
C. Castro et al. hep-th/0107266.
\end{enumerate}
\par I thank my friend and collaborator Sudhir R. Jain for
many discussions.
\end{document}